\documentclass[aps,prb,twocolumn,superscriptaddress]{revtex4}

\usepackage{graphicx}
\usepackage{subfigure}
\usepackage{amsmath}

\def\D{\mathrm{d}}

\begin{document}

\title{Finite-temperature Simulations for Magnetic Nanostructures}
\author{M.A. Novotny}
\affiliation{Center for Computational Sciences, Mississippi State University, Mississippi State, MS, 39762 \texttt{novotny@erc.msstate.edu}}
\affiliation{Department of Physics and Astronomy, Mississippi State University,
Mississippi State, MS, 39762}
\author{D.T. Robb}
\affiliation{ERC Center for Computational Sciences, Mississippi State University, Mississippi State, MS, 39762 \texttt{novotny@erc.msstate.edu}}
\affiliation{Department of Physics and Astronomy, Mississippi State University,
Mississippi State, MS, 39762}
\affiliation{School of Computational Science, Florida State University, Tallahassee, FL, 32306 \texttt{robb@scs.fsu.edu}, \texttt{browngrg@scs.fsu.edu}, \texttt{rikvold@scs.fsu.edu}}
\author{S.M. Stinnett}
\affiliation{ERC Center for Computational Sciences, Mississippi State University, Mississippi State, MS, 39762 \texttt{novotny@erc.msstate.edu}}
\affiliation{Department of Physics and Astronomy, Mississippi State University,
Mississippi State, MS, 39762}
\affiliation{Department of Physics, McNeese State University, Box 93140, Lake Charles, LA, 70609 \texttt{sstinnett@mcneese.edu}}
\author{G. Brown}
\affiliation{School of Computational Science, Florida State University, Tallahassee, FL, 32306 \texttt{robb@scs.fsu.edu}, \texttt{browngrg@scs.fsu.edu}, \texttt{rikvold@scs.fsu.edu}}
\affiliation{Center for Computational Sciences, Oak Ridge National Laboratory, Oak Ridge, TN, 37831}
\author{P.A. Rikvold}
\affiliation{School of Computational Science, Florida State University, Tallahassee, FL, 32306 \texttt{robb@scs.fsu.edu}, \texttt{browngrg@scs.fsu.edu}, \texttt{rikvold@scs.fsu.edu}}
\affiliation{Center for Materials Research and Technology, Florida State University, Tallahassee, FL, 32306}
\affiliation{Department of Physics, Florida State University, Tallahassee, FL 32306}

\begin{abstract}
We examine different models and methods for studying 
finite-tempera-ture magnetic hysteresis in nanoparticles and 
ultrathin films.  This includes micromagnetic results for the 
hysteresis of a single magnetic nanoparticle which is 
misaligned with respect to the magnetic field.  We present 
results from both a representation of the particle as a
one-dimensional array of magnetic rotors, 
and from full micromagnetic simulations.  The results are compared 
with the Stoner-Wohlfarth model.  Results of kinetic Monte Carlo 
simulations of ultrathin films are also presented.  In addition, 
we discuss other topics of current interest in the modeling of 
magnetic hysteresis in nanostructures, 
including kinetic Monte Carlo simulations 
of dynamic phase transitions and First-Order Reversal Curves.
\end{abstract}

\maketitle

%%%%%%%%%%%%%%%%%%%%%%%%%%%%%%%%%%%%%%%%%%%%%%%%%%%%%%%%%%%%%%%
%
%%%%%%%%%%%%%%%%%%%%%%%%%%%%%%%%%%%%%%%%%%%%%%%%%%%%%%%%%%%%%%%
\section{Introduction}

Although hysteresis in fine magnetic particles has been intensively 
studied for many years, there is currently significant 
interest in reexamining our understanding of this phenomenon.  
Partly, this interest is driven by the potential application of 
hysteresis in nanostructures to new technologies such as Magnetic
Random Access Memory (MRAM) and ultra-high-density magnetic recording. 
For the past several years, the areal density of hard drives has been
doubling every 18 months, and is rapidly approaching the limits
of conventional longitudinal recording technology.  At the same time, 
data rates in these drives have increased significantly, with the
serial interface standard at the time of writing 
providing a peak data transfer rate of 2.4 $Gb/s$ \cite{wikkiwikki}.  
This has led the magnetic recording 
industry to look at new recording paradigms such as patterned media and 
self-assembled arrays of nanostructures. 
In fact, the first laptop computer incorporating
a hard drive based on perpendicular recording technology was recently
introduced \cite{EWEEK}. It is crucial, then, to understand the complex
process of hysteresis in these systems.

At the same time, recent advances in computational 
ability, both in terms of new algorithms and 
available computer resources, allow for numerical studies never 
before possible.  Plumer and van Ek~\cite{PLUMER}, for instance, have studied 
the effects of anisotropy distributions in perpendicular 
media using a micromagnetic model.  Their results (Fig. 1) 
show how anisotropy distributions tend to reduce the 
squareness of the loop and, therefore, the signal to noise 
ratio (SNR). Gao et al. have recently carried out similar studies of
tilted perpendicular media \cite{GAO1} and polycrystalline media
\cite{GAO2}.  Another important effect which can be better 
understood through simulations is Barkhausen noise.  This effect 
also decreases SNR, particularly in new thin-film media with 
soft underlayers.  Dahmen, Sethna, and coworkers used a random-field 
Ising model to examine the origins of Barkhausen noise and 
have been able to relate it to avalanches and disorder-induced 
critical behavior~\cite{DAHMEN1,DAHMEN2,DAHMEN3}.  These 
results illustrate two of the ways simulations can be used 
to help understand both fundamental processes in hysteresis 
and their applications to new technology.

Here we present an overview of several common approaches to 
studying hysteresis in magnetic nanostructures.  We then present results
of large-scale computer simulations of hysteresis in single iron nanoparticles 
when the magnetic field is misoriented with respect to 
the long (easy) axis of the elongated particles.  We also 
examine other recent advances in the study of magnetic 
hysteresis, such as kinetic Monte Carlo simulations of dynamic 
phase transitions and First-Order Reversal Curves.
%%%%%%%%%%%%%%%%%%%%%%%%%%%%%%%%%%%%%%%%%%%%%%%%%%%%%%%%%%%%%%%
%  Section II: Models
%%%%%%%%%%%%%%%%%%%%%%%%%%%%%%%%%%%%%%%%%%%%%%%%%%%%%%%%%%%%%%%
\section{Models and Methods}
\subsection{Coherent Rotation}
Given a single-domain particle with uniaxial anisotropy, it is 
possible to find the metastable and stable energy positions of 
the magnetization when a magnetic field is applied at an 
angle to the easy axis. It is assumed that the magnetization 
can be represented by a single vector $\vec{M}$, with constant 
amplitude, $M_{S}$. The energy density of the system is then
\begin{equation}
E=K\sin^{2}\vartheta-M_{S}H\cos(\phi-\vartheta)\;,
\end{equation}
where $K$ is the uniaxial anisotropy constant, $H$ is the 
magnetic field applied at an angle $\phi$ to the easy axis, 
and $\vartheta$ is the angle the magnetization makes with 
the easy axis.  Stoner and Wohlfarth showed that for 
coherent reversal of the magnetization, the spinodal 
curve beyond which the metastable energy minimum 
disappears and switching occurs is given by~\cite{stoner},
\begin{equation}
h_{AX}^{2/3}+h_{AY}^{2/3}=1\;,
\end{equation}
where $h_{AX}$ and $h_{AY}$ are the respective components of the magnetic 
field normalized by the anisotropy field $H_{K}=2K/M_{S}$, along the easy 
and hard axes. Equation (2) is the well-known equation of a hypocycloid 
of four cusps, also known as an astroid.
\index{Stoner--Wohlfarth model}

\begin{figure*}[t]
\begin{center}
\includegraphics[width=0.61\textwidth]{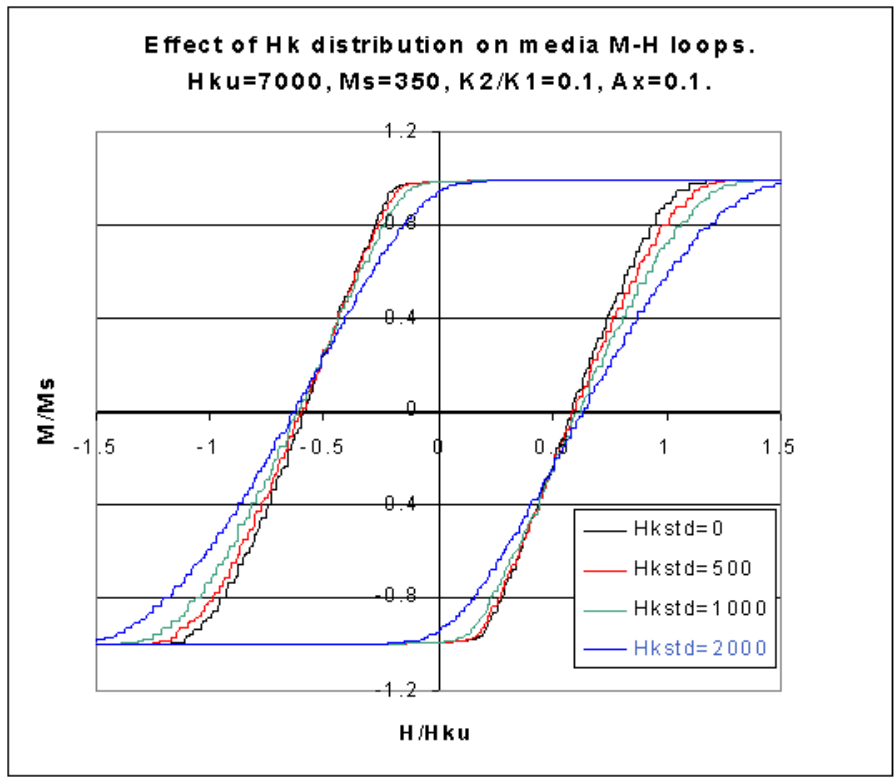}
\end{center}
\caption{The effect of anisotropy distributions on 
hysteresis in perpendicular media.  A Gaussian distribution 
of anisotropy fields is used with a mean value of 
$H_{ku}$~=~7000\,Oe.  In the bottom left corner, the curves
correspond to standard deviations in $H_K$ of (from left to right)
2000\,Oe, 1000\,Oe, 500\,Oe, and 0\,Oe.  The saturation magnetization 
is 350\,emu/cm$^3$.  
Data courtesy of M. Plumer~\cite{PLUMER}}
\label{PLUMERPIC}
\end{figure*}

\subsection{Micromagnetics}
For systems in which the spins are not aligned and/or the 
field is changing too rapidly for the magnetization to reach 
its quasi-static value, it is usually necessary to use a 
non-perturbative technique such as micromagnetics 
to describe the reversal process. The basic approach is to 
divide the system into a coarse-grained set of sites.  
Each site is associated with a position $\vec{r}_{i}$, and 
its magnetization is represented by a single magnetization vector 
$\vec{M}\left(\vec{r}_{i}\right)$, whose norm is the saturation  
magnetization $M_{S}$, 
corresponding to the bulk material (a valid assumption for 
temperatures well below the Curie temperature~\cite{garanin}).  
The time evolution of each spin is given by the 
Landau--Lifshitz--Gilbert (LLG) equation~\cite{brown,nowak,greg1},
\begin{equation}
\frac{\D {\vec{M}(\vec{r}_{i})}}{\D t}=
\gamma_{0}\left(\vec{M}(\vec{r}_{i})\times\left
[\vec{H}(\vec{r}_{i})-\frac{\alpha}{\gamma_{0}M_{S}}
\left(\frac{\D \vec{M}
(\vec{r}_{i})}{\D t}\right) \right] \right)\;,
\end{equation}
where $\vec{H}\left(\vec{r}_{i}\right)$ is the total local 
field at the $i$-th site, $\gamma_{0}$ is the gyromagnetic 
ratio $(1.76\times10^{7}$ rad/Oe\,s),
and $\alpha$ is a dimensionless damping parameter which 
determines the rate of energy dissipation in the system. 
The first term represents the precession of each spin 
around the local field, while the second term drives the 
magnetization to align with the field. The LLG equation can easily 
be rewritten in a form more convenient for numerical
integration~\cite{brown,aharoni}
\begin{multline}
\frac{\D \vec{M}(\vec{r}_i)}{\D t}= \frac{\gamma_{0}}{1+\alpha^{2}} \\
\left( \vec{M}(\vec{r}_{i}) \times \left[ \vec{H}(\vec{r}_{i})- \frac{\alpha}{M_{S}} \left( \vec{M}(\vec{r}_{i})\times \vec{H}(\vec{r}_{i}) \right) \right] \right)
\; .
\end{multline}
For the sign of the undamped precession term, we follow 
the convention of Brown~\cite{brown}.

The total local field, $\vec{H}\left(\vec{r}_{i}\right)$, controls 
the dynamics and contains all of the interactions between 
each site and the rest of the system; it is defined by
\begin{equation}
\vec{H}\left(\vec{r}_{i}\right)=
-\frac{\partial E_i}{\partial\vec{M}\left(\vec{r}_{i}\right)}\;.
\end{equation}
Here, $E_i$ is the free energy of the $i$-th site and the operator
$\partial/\partial\vec{M}\left(\vec{r}_i\right)=$
$\left(\partial/\partial M_x\left(\vec{r}_i\right)\right)\hat{x}+
\left(\partial/\partial M_y\left(\vec{r}_i\right)\right)\hat{y}+
\left(\partial/\partial M_z\left(\vec{r}_i\right)\right)\hat{z}$.
The different 
terms that contribute to $\vec{H}\left(\vec{r}_{i}\right)$ combine 
via linear superposition,
\begin{equation}
\vec{H}(\vec{r}_{i})=
\vec{H}_{Z}(\vec{r}_{i})+\vec{H}_{e}(\vec{r}_{i})+
\vec{H}_{D}(\vec{r}_{i})+\vec{H}_{a}(\vec{r}_{i})+
\vec{H}_{n}(\vec{r}_{i})\;.
\end{equation}
Here, $\vec{H}_{Z}\left(\vec{r}_{i}\right)$ is the externally applied
field (Zeeman term), $\vec{H}_{e}\left(\vec{r}_{i}\right)$ is due to 
exchange interactions, $\vec{H}_{D}\left(\vec{r}_{i}\right)$ is the 
dipole field, $\vec{H}_{a}\left(\vec{r}_{i}\right)$ is the anisotropy 
field (in our simulations taken to be zero), and 
$\vec{H}_{n}\left(\vec{r}_{i}\right)$ is a random field representing 
the effects of thermal noise.

The exchange contribution to the local field represents 
local variations between the alignment of 
$\vec{M}\left(\vec{r}_{i}\right)$ and neighboring sites and can be 
represented by $l_{e}^{2}\nabla^{2}\vec{M}
\left(\vec{r}_{i}\right)$~\cite{aharoni}.  
In our simulations, this is implemented by
\begin{equation}
\vec{H}_{e}(\vec{r}_{i})=
\left(\frac{l_{e}}{\Delta r}\right)^{2}
\left(-n_i\vec{M}(\vec{r}_{i})+\sum_{\left|d\right|
=\Delta r}\vec{M}(\vec{r}_{i}+\vec{d})\right)\;,
\end{equation}
where the summation is over the nearest neighbors of 
$\vec{r}_{i}$, $n_i$ is the number of neighbors of site 
$i$, and the term $n_i\vec{M}\left(\vec{r}_i\right)$ is included so 
that $\vec{H}_e=0$ when all of the spins are aligned.  The exchange 
length, $l_e$, is defined in terms of the 
exchange energy~\cite{arrot}, 
$E_e=-\left(l_e^2/2\right)\int d\vec{rM}\cdot\nabla^2\vec{M}$, in a 
\emph{continuous} system. For our discrete system of finite-sized cells,
this means the magnetization can be viewed as rotating continuously from the
center of one cell to the center of each neighboring cell along the line
joining the two.

At non-zero temperatures, thermal fluctuations contribute 
a term to the local field in the form of a stochastic 
field $\vec{H}_{n}\left(\vec{r}_{i}\right)$, which is assumed to 
fluctuate independently for each spin. The fluctuations
are assumed to be Gaussian, with zero mean and 
(co)variance given by the fluctuation-dissipation 
theorem~\cite{brown,greg1},
\begin{equation}
\left\langle H_{n\mu}(\vec{r}_{i},t)
H_{n\mu^\prime}(\vec{r}_{i}^\prime,t^\prime)\right\rangle = 
\frac{2\alpha k_{B}T}{\gamma_{0}M_{S}V}\delta(t-t^\prime)
\delta_{\mu,\mu^\prime}\delta_{i,i^\prime}\;,
\end{equation}
where $H_{n\mu}\left(\vec{r}_{i}\right)$ indicates one of the Cartesian 
coordinates of $\vec{H}_{n}\left(\vec{r}_{i}\right)$. Here, 
$T$ is the absolute temperature, $k_B$ is Boltzmann's constant,
$V=\left(\Delta r\right)^{3}$ is the discretization 
volume of the numerical integration, and $\delta_{r,r^\prime}$ 
is the Kronecker delta representing the orthogonality of the 
Cartesian components. Although this result was derived for an
isolated particle, recent work by Chubykalo, et al. indicates
that this result will hold for interacting systems as 
well~\cite{CHUBYKALO}. 
In this paper, we present micromagnetic results for two 
different models. The first model is of a nanoparticle with 
dimensions 5.2\,nm$\times$5.2\,nm$\times$88.4\,nm. The 
cross-sectional dimensions are small enough 
($\approx$~2 $l_e$) that the assumption is made that the 
only significant inhomogeneities occur along the long axis 
($z$-direction)~\cite{nowak,boerner2}. The particles
of this model are therefore discretized into a linear 
chain of 17 spins along the long axis of the particle.  
We will call this model the stack-of-spins model.

In this simple model, the local field due to dipole-dipole 
interactions is calculated as~\cite{arrot,boerner2}
\begin{equation}
\vec{H}_{D}(\vec{r}_{i})=
(\Delta r)^3\sum_{j\neq i}\frac{3\hat{\vec{r}}_{ij}
(\hat{\vec{r}}_{ij}\cdot \vec{M}(\vec{r}_j))
-\vec{M}(\vec{r}_j)}{\vec{r}_{ij}^3}\;,
\end{equation}
where $\vec{r}_{ij}$ is the displacement vector from the 
center of cube $i$ to the center of cube $j$, and 
$\hat{\vec{r}}_{ij}$ is the corresponding unit vector.  
The volume factor $(\Delta r)^3$ results from 
integration over the constant magnetization density in 
each cell.
The second model, which we will refer to as the full 
micromagnetic model, simulates a single nanoparticle 
with dimensions 9~nm $\times$ 9~nm $\times$ 150~nm.  
The dimensions were chosen to correspond to arrays of 
nanoparticles fabricated by Wirth, et al.~\cite{wirth1}.
In this model, the system is discretized into 4949 sites 
(7 $\times$ 7 $\times$ 101) on the computational lattice.  
The size of the system makes calculation of dipole 
interactions in the conventional manner (as done for 
the stack-of-spins model) computationally impractical.  
It is therefore necessary to use a more advanced
algorithm to make the simulation tractable.  

The two most popular 
choices are the traditional Fast Fourier Transform (FFT)
and the Fast Multipole Method (FMM)~\cite{greengard}.  
Here, we used the Fast Multipole Method, the exact implementation 
of which is discussed elsewhere~\cite{greg1}, because it
has several advantages over the FFT.  The biggest difference is that
the FMM makes no assumptions about the shape of the underlying lattice,
while the FFT assumes a cubic lattice with periodic boundary conditions.
The consequence of this is that numerical models of systems without 
periodic boundary conditions which use the FFT require empty space 
around the system so that the boundary conditions do not affect the 
calculation.  The FMM requires no such ``padding''.  Furthermore, the FFT 
requires $O\left(N{\rm ln} N \right)$ operations to calculate the 
magnetic scalar potential (from which the dipole field is calculated).  
The FMM algorithm, while it has a larger computational overhead, 
requires only $O\left(N\right)$ operations for the same calculation.  
This means that, while the FFT is a good choice for small cubic lattices, 
the FMM is better for large, incomplete, or irregular lattices. The
public-domain psi-Mag toolset now provides a flexible implementation of the FMM
designed for use on high performance, parallel computers \cite{greg2}.

Material properties in both models were chosen to 
correspond to bulk Fe.  The saturation magnetization 
is 1700 emu/cm$^3$ (kA/m) and $l_e = 2.6$\,nm.  We take the 
damping parameter $\alpha = 0.1$ to correspond to the 
underdamped behavior usually assumed to be present in 
nanoscale magnets.  Although this value is approximately an order
of magnitude larger than the value obtained experimentally
using Ferromagnetic Resonance (FMR), it has been noted that
the FMR value is for small deviations of the magnetization from equilibrium
and is not representative of the large deviations which
occur during reversal~\cite{stinnett1}. In general, care should
be taken in establishing an appropriate damping parameter to use in
simulating a particular magnetic nanostructure, as also illustrated
in other recent micromagnetic studies \cite{fidler,hertel,boerner1}.
 
It is worth noting that, even in systems which reverse 
coherently at high speed, deviations from the 
quasi-static Stoner--Wohlfarth (SW) model will be expected.  
He, et al.~\cite{He} showed that for square-pulse fields 
with fast rise times ($<$ 10\,ns) and small values of the 
damping constant ($<$ 0.2), the shape of the astroid changes.  
The result is that the minimum switching field is reduced 
below the SW limit of 0.5 $H_K$, and the angular dependence 
is no longer symmetric around 45$^\circ$.  For the frequencies 
and damping parameter used here, the deviations from 
the SW model are small ($<$ 5 percent difference in the 
switching fields) and may be neglected.
\subsection{Monte Carlo Simulations: Kinetic Ising and Heisenberg Models}
A second approach to modeling the dynamics of magnetic systems
involves Monte Carlo techniques, which have been applied to a wide 
variety of systems since their introduction 
by Metropolis, et al.~\cite{metropolis}.  
As described above, the micromagnetics approach uses a stochastic
(i.e. random number-based) variable to introduce random fluctuations into
an otherwise deterministic system.
In contrast, Monte Carlo simulations are fully stochastic and
proceed by considering possible
transitions between states of the system and
executing these transitions with a probability which depends on
the system's energy and temperature.

The static Monte Carlo algorithm consists of a repeated 
three-step process.  First, choose a (pseudo-)random number.  
The random numbers chosen may be uniformly distributed, 
or they may be chosen based on a particular probability 
distribution that depends on the 
specific simulation to be performed.  Second, choose a 
trial move from the current state to a new state.  Third, 
accept or reject the trial move depending on the random 
number and some acceptance rule consistent with the problem 
under consideration.

Consider the ferromagnetic Ising model on a regular lattice 
with periodic boundary conditions.  Each site on the lattice 
has a spin which can align either parallel or anti-parallel 
to the applied field and takes on values of $S_i=\pm 1$ 
accordingly.  The energy of the Ising lattice is then
\begin{equation}
E=-J\sum_{\langle i,j\rangle}S_{i}S_{j}-H(t)\sum_{i}S_i\;,
\end{equation}
where the exchange constant $J>0$ is in units of energy,
and $H(t)$ is the externally applied, time-dependent 
magnetic field (which in Monte Carlo simulations is customarily
given in units of energy, thus absorbing the magnetic moment 
per site, $\mu$).  The first sum in (10) is over nearest-neighbor 
pairs, while the second sum is over all spins on the 
lattice.
  
The static Monte Carlo procedure described above 
allows the calculation of equilibrium quantities such as 
the internal energy, susceptibility, specific heat, and 
magnetization.  In order for the lattice to explore each
of its possible states with probabilities corresponding 
to the equilibrium thermal distribution, the acceptance 
rule chosen must satisfy the condition of detailed balance~\cite{Landau}.
Two common choices are the Metropolis~\cite{metropolis} and 
Glauber~\cite{GLAUBER} acceptance rules.  Note that near the 
critical point, computation with these simple acceptance 
rules slows down dramatically, and it is therefore useful 
to use more advanced algorithms (such as cluster 
algorithms~\cite{Landau}) to calculate equilibrium 
quantities.

In equilibrium calculations, no physical interpretation is 
ascribed to the intermediate spin flips.  If, instead, we 
consider the individual spin flips as representing physical
fluctuations due to the interactions 
between the spins and a heat bath, then the underlying 
transitions model the actual dynamics of the system and 
acquire a physical significance.  This application
of Monte Carlo simulations is known as kinetic Monte 
Carlo.  The random nature of the events due to 
the interaction of spins dictates 
that the spin to attempt to flip must be chosen at random.
In this paper, we use the Glauber~\cite{GLAUBER} 
acceptance rule, according to which each attempted spin 
flip is accepted with probability
\begin{equation}
W~=~\frac{\exp\left( -\beta\Delta E_i\right)}
{1+\exp\left( -\beta\Delta E_i\right)}\;.
\end{equation}
Here, $\Delta E_i$ is the change in energy that results
if the proposed flip of the $i$-th spin is accepted, and 
$\beta$~=~$\left( k_BT\right)^{-1}$.
With a uniformly distributed random number, $r\in [0,1]$, 
a randomly chosen spin is flipped if $r~\leq W$.  Each 
potential spin flip is considered a Monte Carlo step.  The
basic time step of the Monte Carlo process is measured in
Monte Carlo Steps per Site (MCSS).  This time is related
to the algorithm and in general is only approximately 
proportional to the physical time of the system. Recently,
however, progress has been made in connecting analytically
the MC simulation time to the simulation time of the Langevin-based
micromagnetic techniques discussed above, for which there is a
clear relationship to physical time ~\cite{nowak2,chubykalo2,cheng1,cheng2}.

The Glauber 
dynamic of (11) can be derived from a quantum 
spin-$\frac{1}{2}$ Hamiltonian coupled to a fermionic heat 
bath~\cite{MARTIN}.  Recently, other dynamics have been 
derived from coupling a quantum spin-$\frac{1}{2}$ system to a 
phonon heat bath~\cite{PARK}.  Note that in kinetic Monte 
Carlo calculations, algorithms (such as the cluster 
algorithm) that change the underlying dynamic cannot 
be used.  However, advanced algorithms that achieve very
large speedups while remaining 
true to the underlying dynamics are possible~\cite{novotny}.
It has recently been shown that physically relevant functional
forms for $W$ can lead to dramatically different values of 
dynamical quantities such as lifetimes of metastable states~\cite{NEWPRL}.

%%%%%%%%%%%%%%%%%%%%%%%%%%%%%%%%%%%%%%
%\subsection{Classical Heisenberg model}
%%%%%%%%%%%%%%%%%%%%%%%%%%%%%%%%%%%%%
\index{Heisenberg model}
\index{Monte Carlo}

It is important to realize that the Monte Carlo techniques 
can be applied to other systems as well.
Unlike the Ising model, 
the Heisenberg model allows the spins 
to assume any angle with respect to neighboring 
spins and the applied field.  The energy of a 
regular lattice of Heisenberg spins with periodic 
boundary conditions is
\begin{equation}
E=-J\sum_{\langle i,j\rangle}(S_{ix}S_{jx}+S_{iy}S_{jy}
+S_{iz}S_{jz})-H\sum_{i}S_i\cos (\theta_i)\;,
\end{equation}
where $S_{ix}$, $S_{iy}$, and $S_{iz}$ are the 
Cartesian coordinates of the vector spin $\textbf{S}_i$ 
(with magnitude $S_i = 1$), and $\theta_i$ is the angle between 
the applied field, $H$, and the $i$-th spin. As in the Ising model, 
the first sum is taken
over nearest neighbors and represents the exchange 
interactions, while the second sum is taken over all spins in the system, 
and represents the interactions of the spins with an externally 
applied magnetic field (Zeeman energy).

The dynamic consists of randomly choosing a spin to update, randomly 
choosing a new spin direction (either uniformly distributed over 
the sphere or over a cone near the current spin direction),
and using a Metropolis or a heat-bath rate to decide whether
to effect a transition to the new spin direction. The rate depends 
on the energy difference between the spin configurations
as in, for example, (11).

In kinetic Monte Carlo, it is possible to implement the algorithm
in a rejection-free manner, so that every algorithmic step performs an
update. In this case, each algorithmic step in general advances
the system by a different amount of time. For models, such as the Ising model,
with discrete state spaces this is called the 
$n$-fold way algorithm~\cite{bortz}.
It is possible to make a precise connection between these the $n$-fold way
and the standard implementation of kinetic Monte Carlo~\cite{novotny}. 
Recently such rejection-free methods have been implemented for models with 
continuous state spaces, such as the Heisenberg model~\cite{MUNOZ}, and the
efficiency of rejection-free methods in various systems has been studied
\cite{watanabe}. 

\index{Heisenberg model}

%%%%%%%%%%%%%%%%%%%%%%%%%%%%%%%%%%%%%%%%%%%%%%%%%%%%%%%%%%%%
%  Section III Simulation Results
%%%%%%%%%%%%%%%%%%%%%%%%%%%%%%%%%%%%%%%%%%%%%%%%%%%%%%%%%%%%
\section{Results of Micromagnetic Simulations}
\index{micromagnetics}
\index{hysteresis}
In this section we summarize recent simulation results for
magnetization reversal in iron nanopillars \cite{greg4}, and further 
evaluate these results in light of additional experimental data on such 
reversal.

\begin{figure*}
\vspace{0.2in}
\begin{center}
\mbox{
\subfigure[Hysteresis loops for the full micromagnetic model 
       at $T$~$=$~100\,K with a sinusoidal field of period 
       25\,ns and maximum applied field of 10\,kOe \newline \newline]
 {\includegraphics[width=3.0in, height=2.0in]{figure2a.eps}}}
\mbox{
\subfigure[Hysteresis loops for the stack-of-spins model 
       at $T$~$=$~10\,K with a sinusoidal field of period 
       200\,ns and a maximum applied field of 5\,kOe]
 {\includegraphics[width=3.0in, height=2.0in]{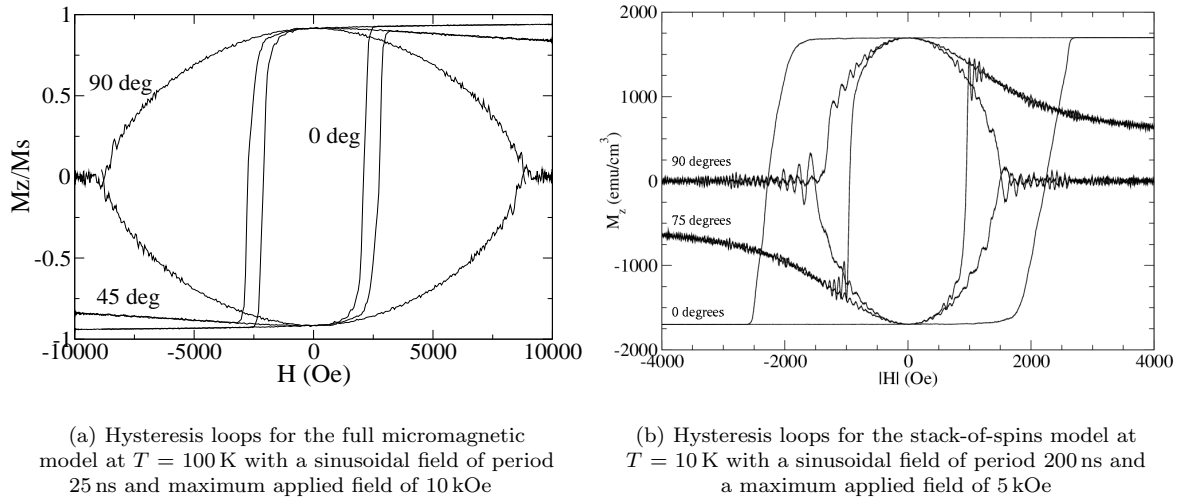}}}
\end{center}
\caption[]{Hysteresis loops for ({\bf a}) the full micromagnetic model with 
       0$^\circ$, 45$^\circ$, and 90$^\circ$ misalignment 
       and ({\bf b}) the stack-of-spins model with 0$^\circ$, 75$^\circ$, 
       and 90$^\circ$ between the applied field and the long 
       axis of the particle}
\label{loops}
\end{figure*}

Figure~\ref{loops}a shows hysteresis loops at $T$~$=$~100\,K 
for the full micromagnetic model with the field misaligned 
at $0^\circ$, $45^\circ$, and $90^\circ$ to the long axis 
of the particle.  The loops were calculated using a sinusoidal 
field with a period of 25\,ns, which started at a maximum 
value of 10,000\,Oe (800 kA/m).  In all the loops in this 
section, the reported 
magnetization is the component along the long axis ($z$-axis) 
of the particle. Simulations for the full micromagnetic model were 
performed over one half of the period and the results 
reflected to give the full hysteresis loop.

Consider the case with the field and particle aligned
($0^\circ$).  Initially, the large magnetic field tends 
to align the spins with the easy axis.  As the field is 
decreased, the spins relax, and the magnetization
decreases by approximately $2\%$.  Eventually, reversal 
initiates at the ends as previously reported~\cite{greg1}.

\begin{figure*}[t]
\mbox{\subfigure[Snapshots of reversal 
                 at times (left to right) $t$~$=$~$7.350$, 
                 $7.375$, $7.400$, $7.425$, and $7.450$\,ns]
 {\includegraphics[width=0.48\textwidth]{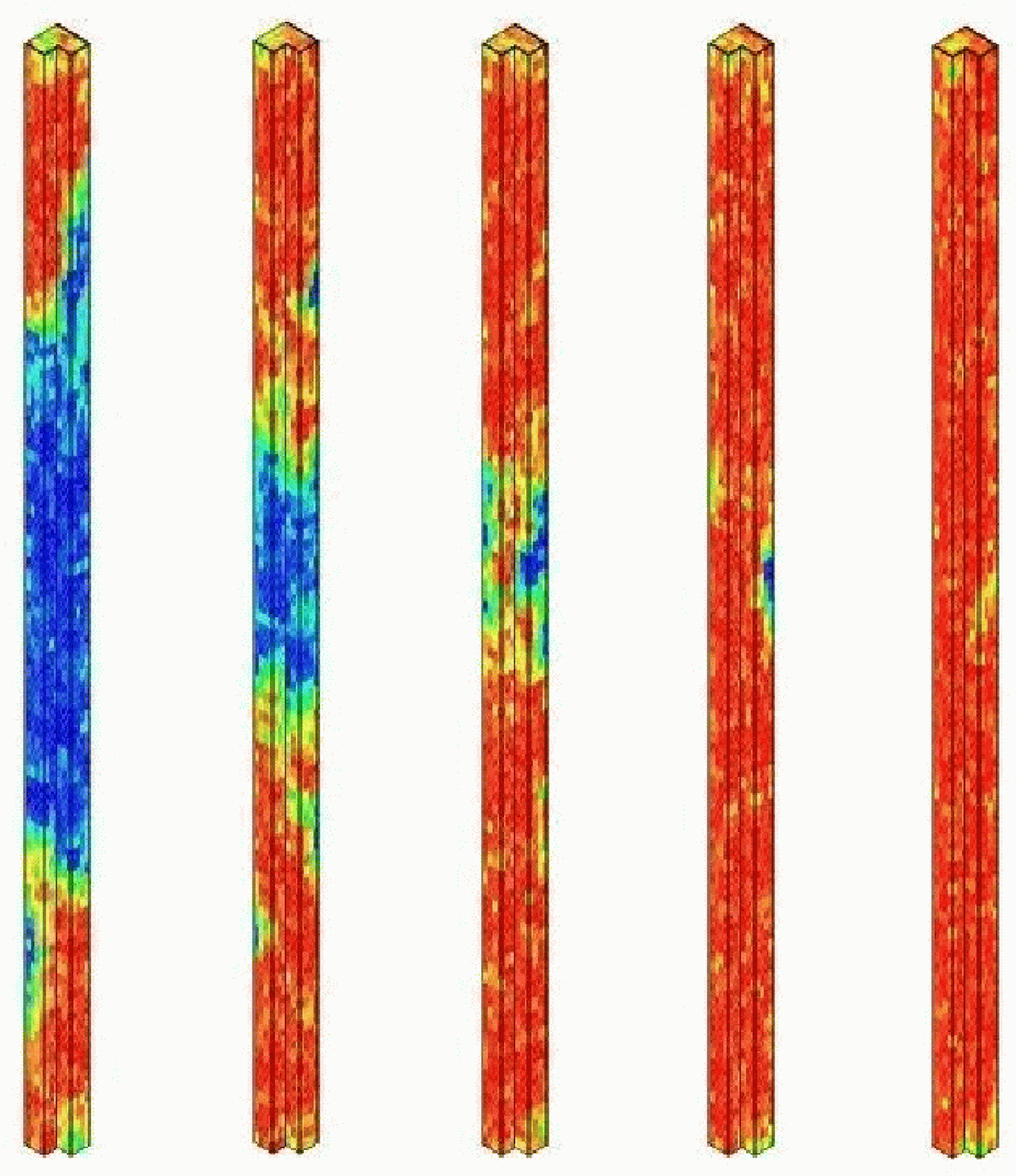}}
      \subfigure[Snapshots of reversal at times 
                 (left to right) $t$~$=$~$0.00$, $3.000$, 
                 $4.000$, $6.250$, and $8.500$\,ns]
 {\includegraphics[width=0.48\textwidth]{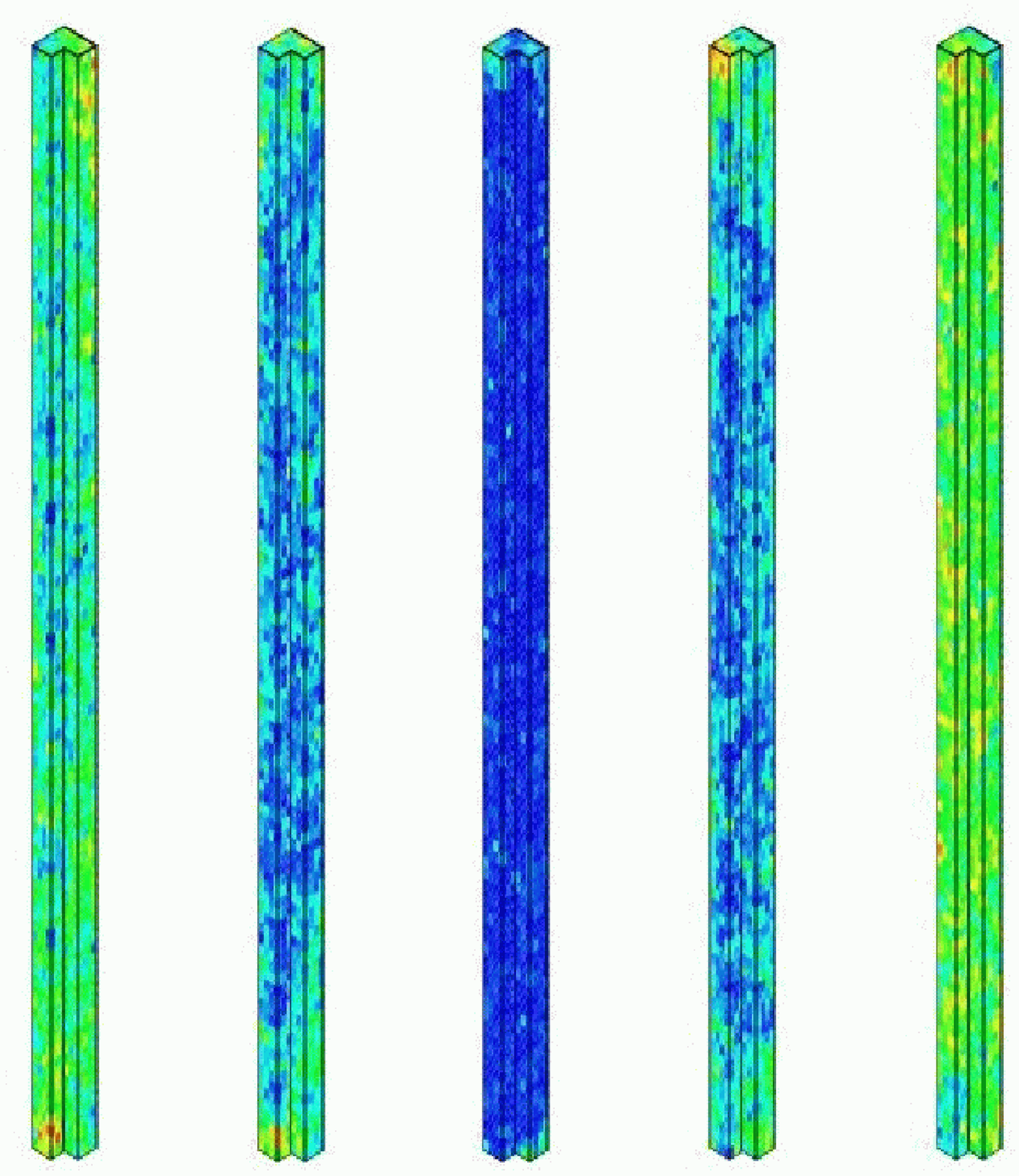}}}
\caption[]{The $z$-component of the magnetization 
           in the full micromagnetic model for the 
           ({\bf a}) 45$^\circ$ and ({\bf b}) 90$^\circ$ hysteresis 
           loops of Fig. 1a}
\label{pics}
\end{figure*}

At $45^\circ$ misalignment between the particle 
and the field, the magnetization is initially pulled away 
from the long (easy) axis by the large magnetic field.  As 
the field is swept toward zero, the magnetization relaxes 
until it essentially reaches a maximum value of 
approximately $0.91 M_S$ at zero applied field.  Thermal 
fluctuations along the length of the particle prevent the 
magnetization from reaching saturation.  As in the case of 0$^\circ$, 
reversal again begins by nucleation at the ends of the 
particle, with the growth of these nucleated regions 
leading to the reversal of the particle.  Figure~\ref{pics}a 
shows the $z$-component of the magnetization at selected times 
during the reversal process for the $45^\circ$ hysteresis loop 
of Fig.~\ref{loops}a. It is important to note that 
the particles do not have a uniform magnetization during the
reversal process, even though they are single-domain particles.

For $90^\circ$ misalignment, the reversal mechanism is quite 
different.  The hysteresis loop in Fig.~\ref{loops}a 
shows that the magnetization is essentially perpendicular 
to the easy direction until the field reaches a particular 
value.  As the field is decreased further, the magnetization 
relaxes toward the easy axis.  Since nothing breaks the up/down 
symmetry of the system when the applied field has no component 
along the easy axis, the relaxed magnetization can be 
directed toward either the positive or negative $z$-axis. 
Figure~\ref{pics}b shows the $z$-component of the magnetization 
for the $90^\circ$ misalignment at selected times during the 
hysteresis loop of the full micromagnetic model. For this 
case, the nucleation occurs along the entire length of the 
particle, except at the ends. The large demagnetizing fields 
present at the ends (involved in nucleation at smaller angles) 
retard relaxation along the easy axis.

The hysteresis loops for the stack-of-spins model, 
shown in Fig.~\ref{loops}b, are qualitatively 
similar to those of Fig.~\ref{loops}a.  Loops 
at $0^\circ$, $75^\circ$, and $90^\circ$ 
misalignment are shown.  There are important 
differences between the two models, however.  
First, without lateral resolution of the 
magnetization across the cross-section, these 
particles exhibit ringing due to the precessional 
dynamics. Evidently, the precession of individual 
moments in the full micromagnetic model does not 
lead to precession of the end-cap moment; possibly 
the spin waves rapidly damp out the gyromagnetic 
motion.

\begin{figure*}[t]
\begin{center}
\includegraphics[width=0.55\textwidth]{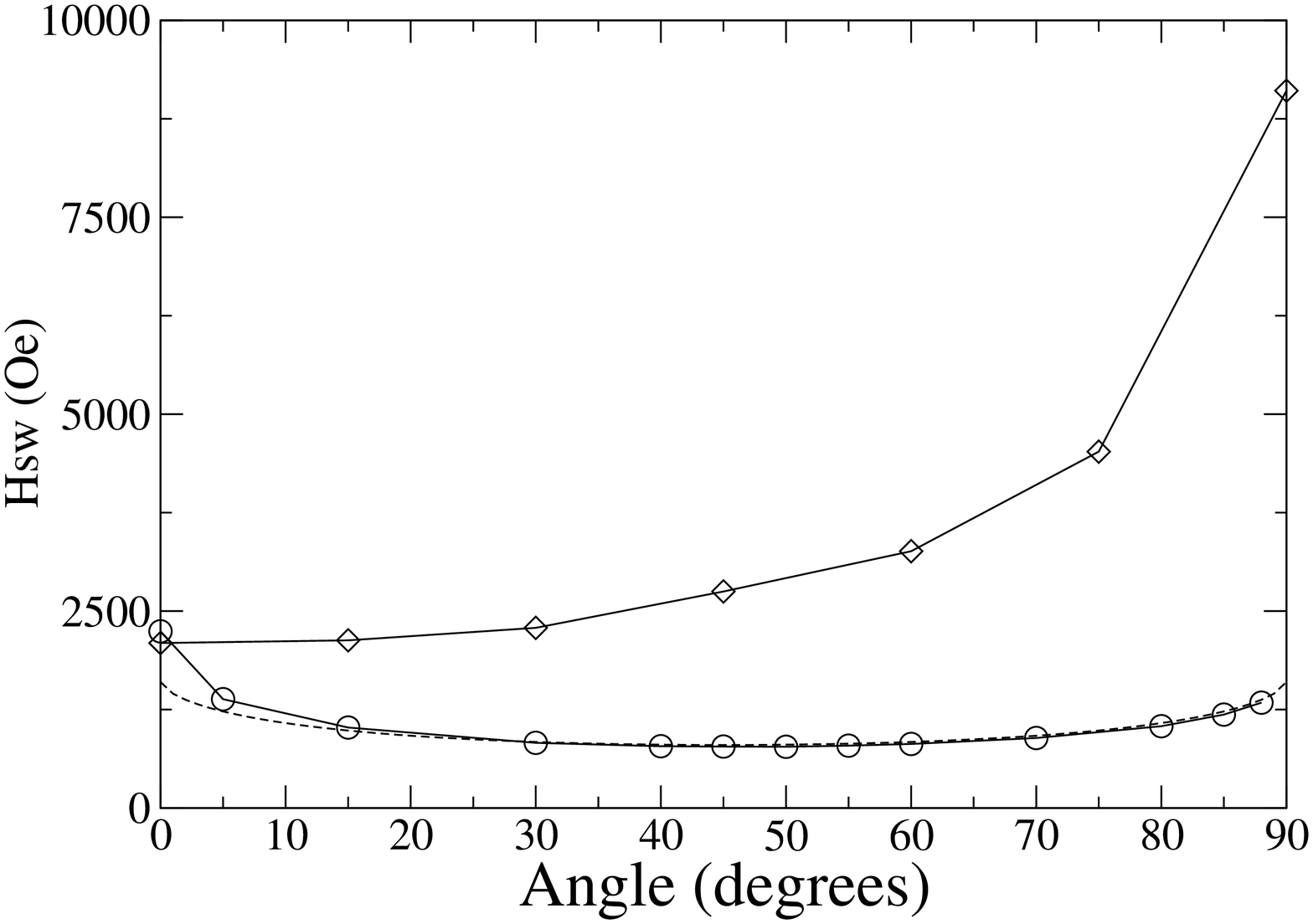}
\end{center}
\caption[]{Angular dependence of the switching 
           field for three models of magnetization 
           reversal. The full micromagnetic model 
           ({\it diamonds}) at $T$~$=$~100\,K shows a 
           distinctly different behavior from both 
           the stack-of-spins model at $T$~$=$~20\,K 
           ({\it circles}) and the Stoner--Wohlfarth 
           model ({\it dashed line}) \newline \newline}
\label{angdep1}
\end{figure*}

\begin{figure*}
\vspace{0.4cm}
\mbox{
\subfigure[Applied field periods are 15\,ns ({\it circles}), 
           25\,ns ({\it squares}), 50\,ns ({\it diamonds}), and 
           100\,ns ({\it triangles}).  The field is sinusoidal 
           with a maximum value of 5\,kOe]
{\includegraphics[width=0.50\textwidth]{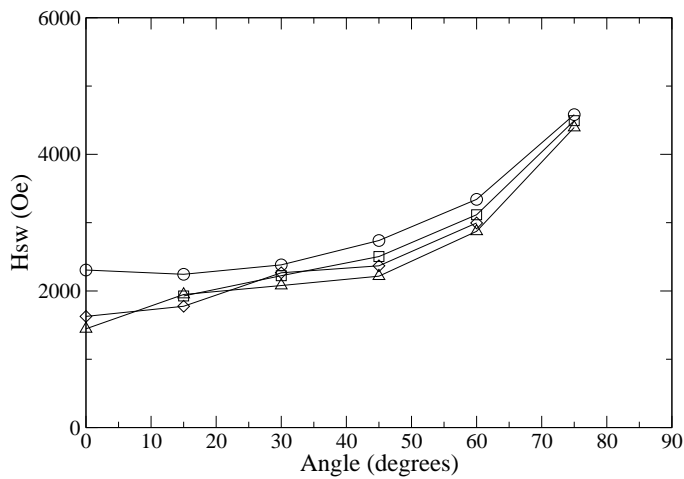}}
\subfigure[Applied field periods are 15\,ns ({\it squares}) 
           and 25\,ns ({\it circles}).  The field is sinusoidal 
           with a maximum value of 10\,kOe]
{\includegraphics[width=0.50\textwidth]{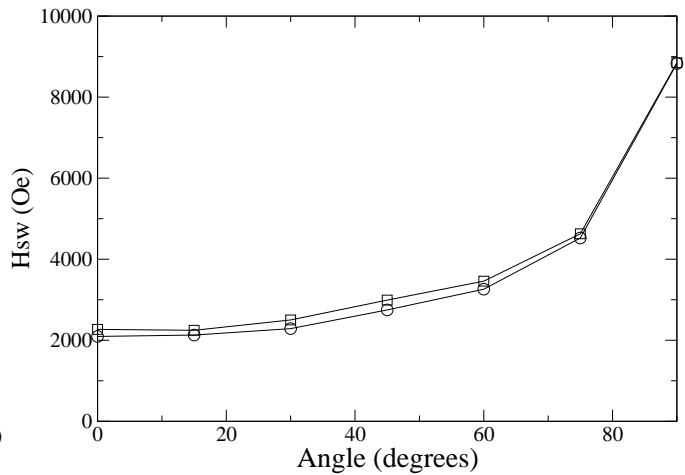}}}
\caption[]{Angular dependence of the switching field for the 
           full micromagnetic model at ({\bf a}) 0\,K and ({\bf b}) 100\,K.  
           At 0\,K, the LLG equation is completely deterministic, 
           while at 100\,K, it includes random fluctuations through 
           a stochastic thermal field.}
\label{angdep2}
\end{figure*}

A second, and more prominent, difference between 
the models is observed in the angular dependence 
of the switching field, $H_{{\rm sw}}$, shown in 
Fig.~\ref{angdep1}.  Here, $H_{{\rm sw}}$
is defined as the applied field at which $M_z$ is 
reduced to 0.  The stack-of-spins model (circles) shows 
a shape qualitatively similar to what is expected 
from Stoner--Wohlfarth (SW) theory, with a minimum 
$H_{{\rm sw}}$ near $45^\circ.$  The dashed curve 
is the SW theory with $H_K$~$=$~1600\,Oe (for 
comparison reasons $H_K$ was chosen to be much smaller 
than the $10^4$\,Oe expected for these particles 
assuming SW behavior). The full micromagnetic model 
(diamonds), on the other hand, has its minimum 
$H_{{\rm sw}}$ at $0^\circ,$ and $H_{{\rm sw}}$ 
increases as the misalignment angle is increased.  
Figure~\ref{angdep2} shows the angular dependence of 
the switching field for the full micromagnetic model 
for periods of 15, 25, 50, and 100\,ns and maximum 
applied field of 5\,kOe at $T$~$=$~0\,K, and for periods of 
15\,ns and 25\,ns with a maximum applied field of 
10\,kOe at $T$~$=$~100\,K.  At 0\,K, the general trend 
is for longer periods to reduce the switching field.  
However, at $15^\circ$, the 100\,ns loop is observed to switch 
at a lower field than either the 50 or 25\,ns loops.  Similarly, 
at $30^\circ$, the 50\,ns loop switches at a lower field than 
the 25\,ns loop.  One reason for this may be
resonance in the switching fields for these angles and 
periods.  At $T$~$=$~100\,K, the 25\,ns loops switch at a lower field 
than the 15\,ns loops for all angles.  At $90^\circ$, where
thermal fluctuations are most prominent, the field at 
which relaxation occurs is independent
of the period within the accuracy of the simulation.

The increase of $H_{\rm{sw}}$ with the misalignment angle in the
micromagnetic simulation is consistent with recent experimental 
observations of Fe nanopillars~\cite{wirth1,wirth2,Li1,Li2,Li3}.
However, the most recent experiment~\cite{Li3} shows
that a nanopillar with lateral dimension $d\sim 5.2~\mathrm{nm}$, which
our formulation suggests should show a dependence of $H_{\rm{sw}}$ on 
misalignment angle similiar to the stack-of-spins model 
(i.e. like coherent rotation), actually exhibits the increasing dependence 
found in the full micromagnetic model. In addition, a nanopillar with lateral
dimension $d\sim 10-15~\mathrm{nm}$ showed evidence of a multi-domain
remanence state. As noted in Ref.~\onlinecite{Li3}, imperfections of the
nanopillar structure appear to contribute to localized nucleation processes
down to smaller than expected lateral dimensions, and probably also provide
the pinning sites causing the multi-domain remanence state. 
This illustrates the importance of coordinating experimental 
and simulation results in the micromagnetic approach. Further
improvement of the predictions of the micromagnetic approach will likely
have to incorporate such structural imperfections.

%%%%%%%%%%%%%%%%%%%%%%%%%%%%%%%%%%%%%%%%%%%%%%%%%%%%%%%
% Section IV: Other Recent Topics in Hysteresis
%%%%%%%%%%%%%%%%%%%%%%%%%%%%%%%%%%%%%%%%%%%%%%%%%%%%%%%
\section{Recent Results for the 2D Kinetic Ising Model}
Monte Carlo simulation of the Ising model, as well as other magnetic 
systems, continues to be an active field of research.
Here, we present three recent results that are of interest in
understanding the process of magnetization reversal in ultra-thin films.
\subsection{Dynamic Phase Transitions}
\begin{figure*}[t]
\mbox{\subfigure[]
 {\includegraphics[angle=270,width=0.50\textwidth]{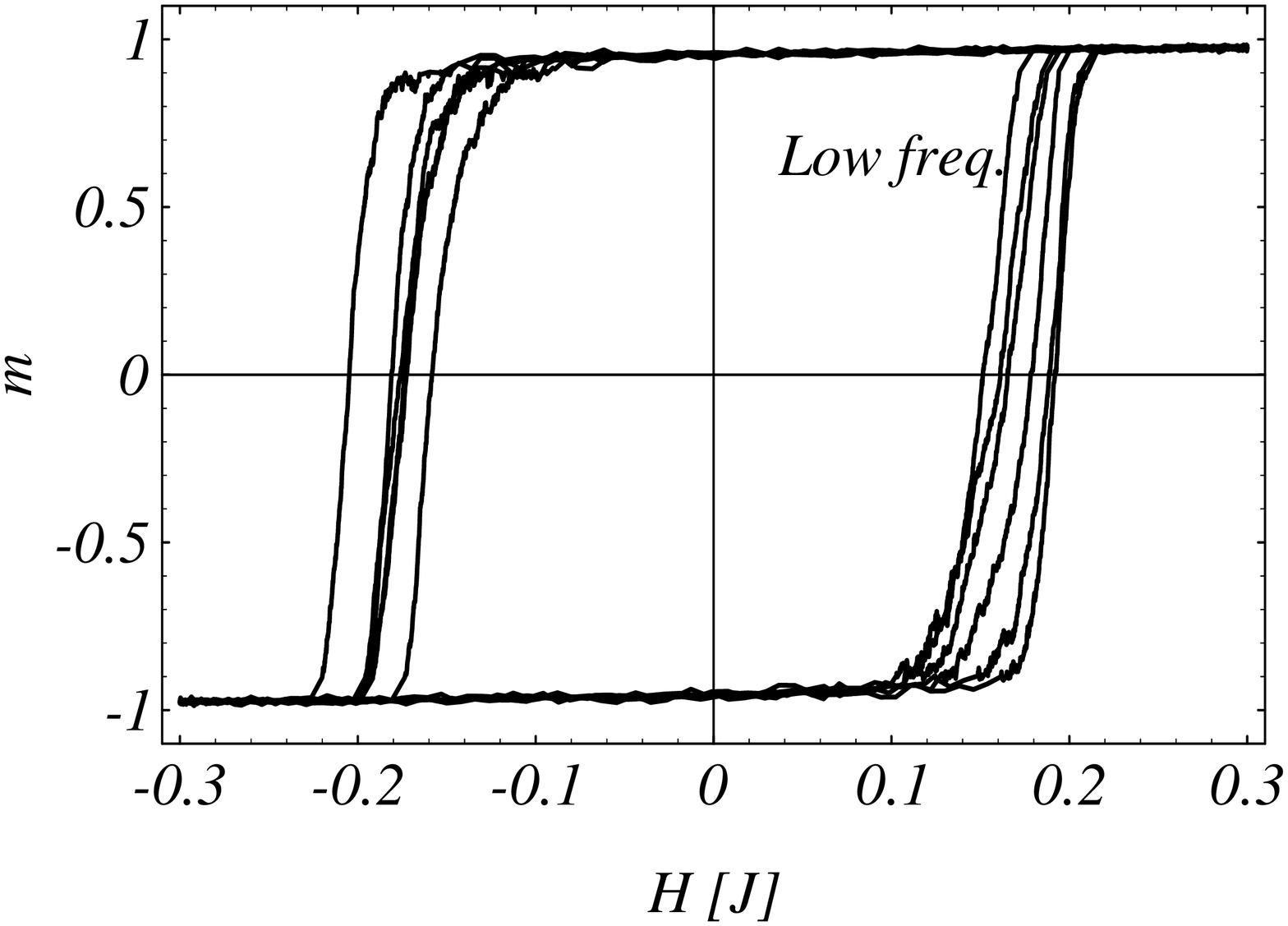}}
      \subfigure[]
 {\includegraphics[angle=270,width=0.50\textwidth]{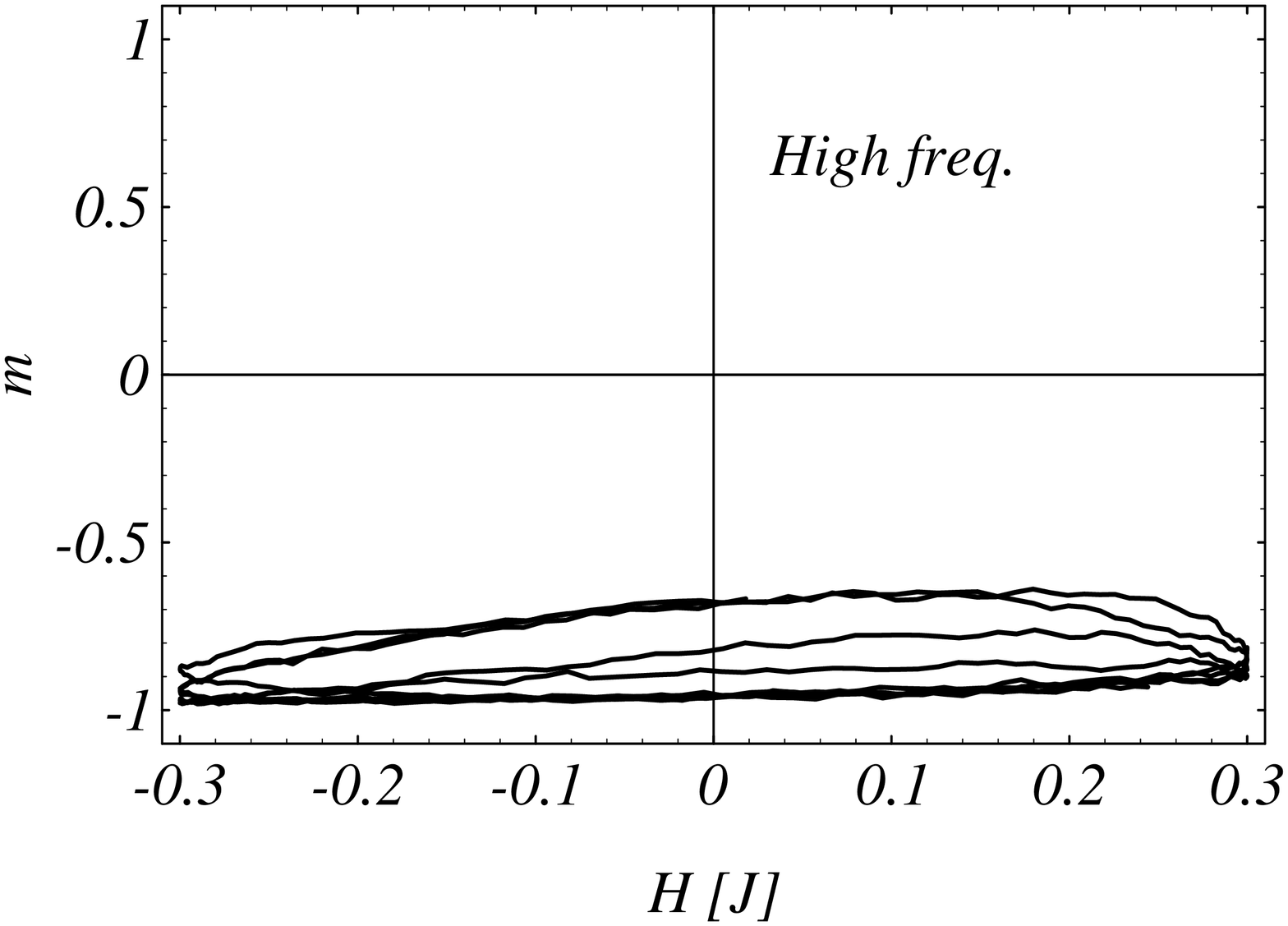}}}
\caption[]{Simulated hysteresis loops for a kinetic Ising model 
           ({\bf a}) in the dynamically disordered phase 
               for $\varTheta\gg\varTheta_c$ and  
           ({\bf b}) in the dynamically ordered phase for 
               $\varTheta\ll\varTheta_c$.  
               Data courtesy of S.W. Sides}
\label{PERLOOPS}
\end{figure*}

\index{micromagnetics}
\index{hysteresis}
\index{dynamic phase transition}
When the half-period $t_{1/2}$ of the applied field is 
longer than the characteristic switching time in a 
constant field, $\langle\tau(H_0)\rangle$, where $H_{0}$ 
is the amplitude of the oscillating field, the 
magnetization can follow the changing field, resulting 
in standard hysteresis loops, such as those shown in 
Fig.~\ref{PLUMERPIC} and in Fig.~\ref{PERLOOPS}a.  However, when 
$t_{1/2}\ll\langle\tau(H_0)\rangle$, the magnetization 
cannot follow the field, but rather oscillates around 
one or the other of its zero-field stable values. This 
breaking of the symmetry of the hysteresis loop is 
associated with a dynamic phase transition (DPT) located 
at an intermediate value of the half-period. In terms of 
the dimensionless half-period, 
$\varTheta=t_{1/2}/\langle\tau(H_0)\rangle$, the transition 
is located at $\Theta_c\approx1$. The dynamic order 
parameter for this transition is the period-averaged 
magnetization,
\begin{equation}
Q_n=\frac{1}{2t_{1/2}}\int_{(n-1)(2t_{1/2})}^{n(2t_{1/2})} {m(t)\D t}\;.
\end{equation}

\begin{figure*}
\begin{center}
\includegraphics[width=0.5\textwidth]{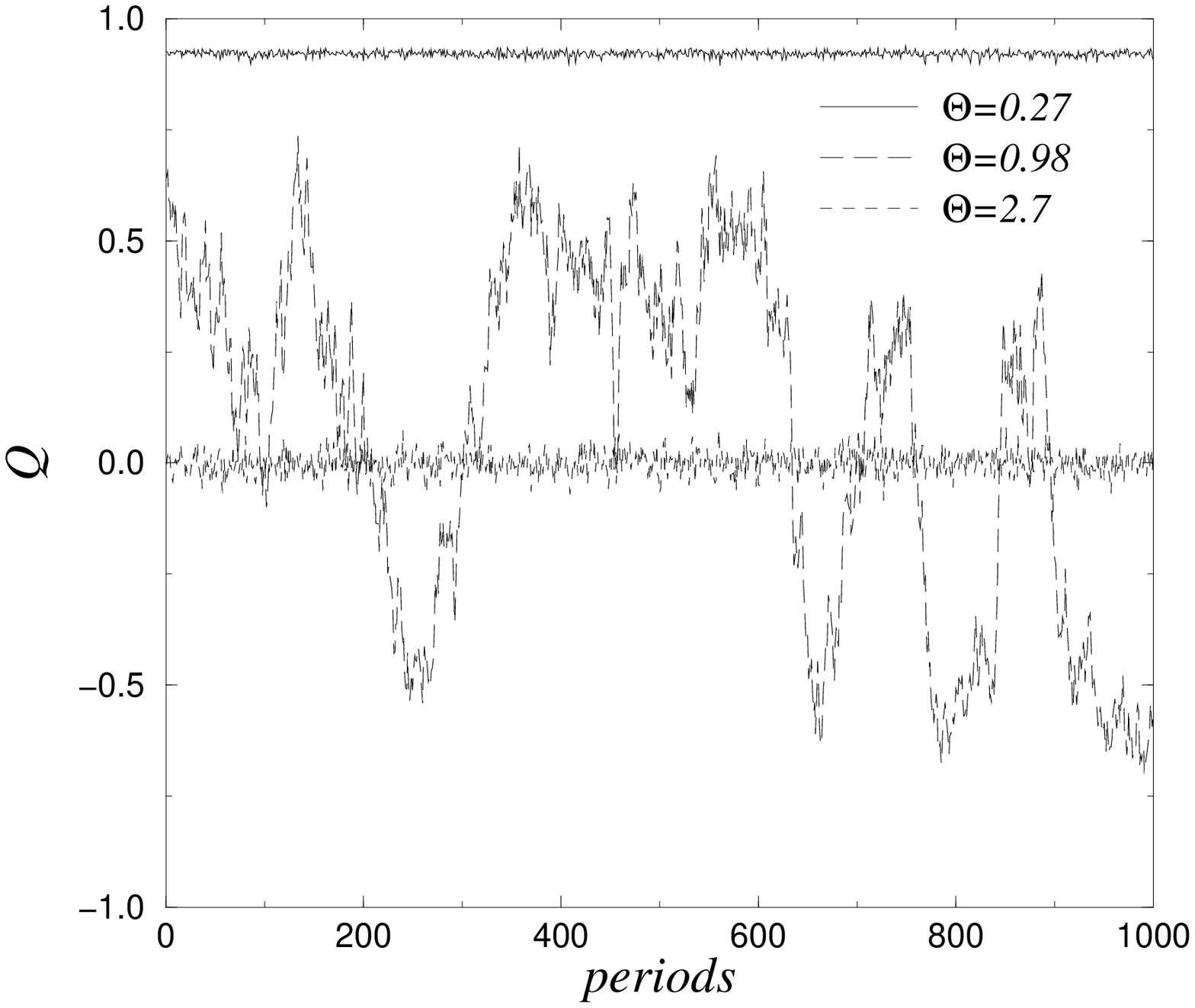}
\end{center}
\caption[]{Time series of the dynamic order parameter 
           $Q_n$ in the dynamically ordered phase 
           (curve near +1, $\varTheta$~$=$~0.27), near the transition 
           (curve fluctuating wildly about zero, $\varTheta$~$=$~0.98), 
           and in the dynamically disordered phase 
           (curve that remains close to zero, $\varTheta$~=~2.7). 
           After ~\cite{ISING3}}
\label{PER2}
\end{figure*}

In Fig.~\ref{PERLOOPS}, we show hysteresis loops for the two-dimensional 
kinetic Ising model using Glauber dynamics for the 
dynamically disordered phase with $\varTheta\gg\varTheta_c$ 
and the dynamically ordered phase with $\varTheta\ll\varTheta_c$. 
Time series of $Q_n$ for $\varTheta\gg\varTheta_c$, 
$\varTheta\approx\varTheta_c$, and $\varTheta\ll\varTheta_c$ are 
shown in Fig.~\ref{PER2}.

The DPT was first discovered in numerical solutions 
of a mean-field model of a ferromagnet in an 
oscillating field~\cite{TOME,MENDES}. It has since 
been intensively studied in mean-field 
models~\cite{MFA1,MFA2,MFA3}, kinetic Ising 
models ~\cite{ISING1A,ISING1B,ISING1C,ISING2,ISING3,ISING4}, 
the kinetic spherical model ~\cite{Paessens}, and
anisotropic $XY$ ~\cite{TUTU,fujiwara} and Heisenberg 
\cite{JANG,huang} models.  There have also been indications 
of its presence in experimental studies of hysteresis in 
ultra-thin films of Cu on Co(001)~\cite{EXP,EXP2}. 
From a theoretical point of view, its most interesting 
feature is that this {\it far-from-equilibrium} phase 
transition is a genuine continuous (second-order) phase 
transition that belongs to the {\it same} universality 
class as the {\it equilibrium} phase transition in the 
Ising model in zero field~\cite{ISING2,ISING3,ISING4,FUJI}.  
Unequivocal experimental verification of this interesting 
non-equilibrium phase transition is highly desirable and, given
that new high-density magnetic recording media will require shorter
reversal periods, may be relevant to the design of magnetic 
storage devices.
\subsection{Hysteresis Loop Area and Stochastic Resonance}
\index{stochastic resonance}

\begin{figure*}[t]
\begin{center}
\includegraphics[width=0.63\textwidth]{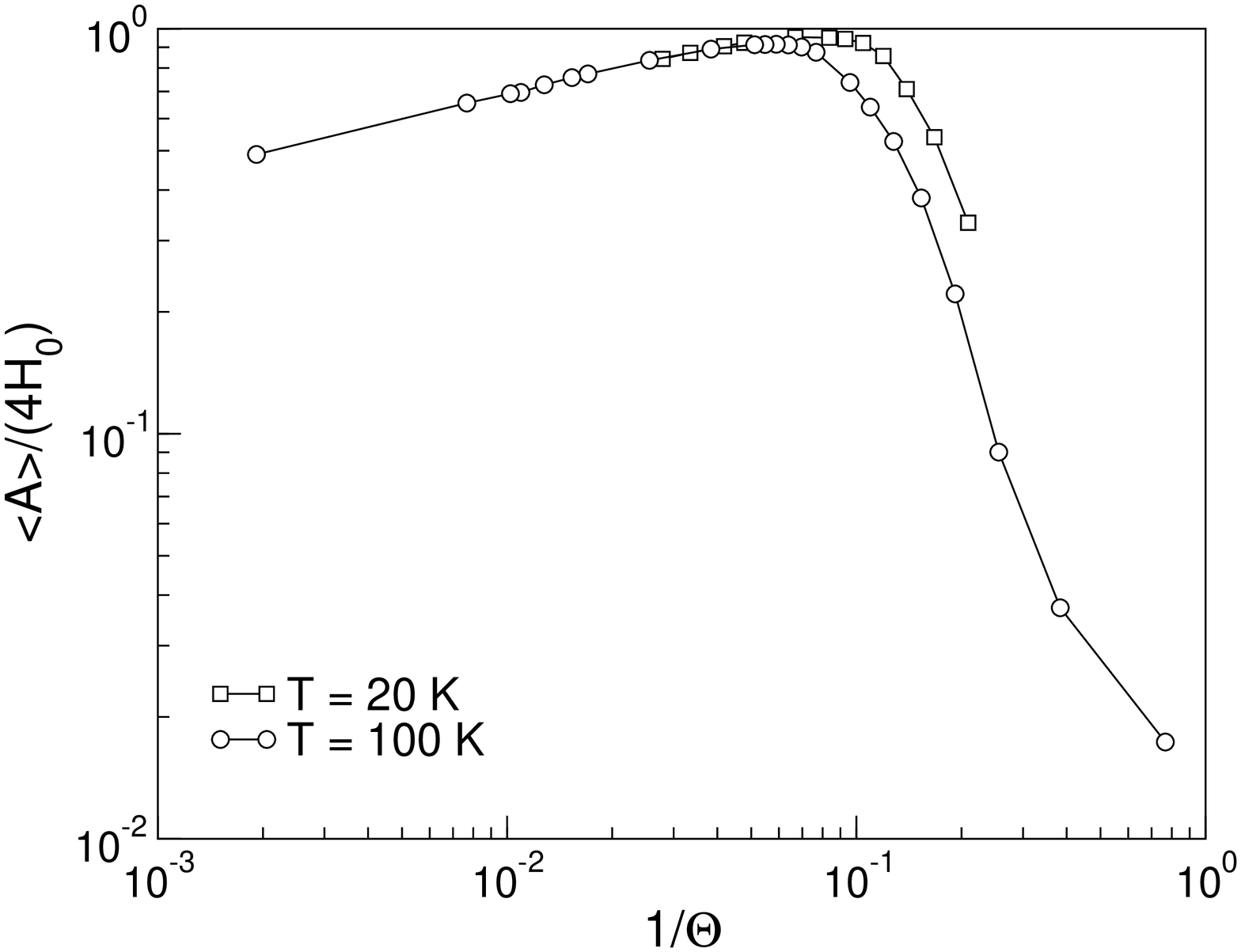}
\end{center}
\caption[]{Average hysteresis-loop area, $\langle A\rangle$, 
           vs scaled frequency, $1/\varTheta$ for the stack-of-spins 
           micromagnetic model.  The same behavior is seen in 
           two-dimensional Ising models that switch by a 
           single-droplet mechanism, and the maximum is associated 
           with stochastic energy resonance}
\label{areas}
\end{figure*}

The values of $Q$ measured for the stack-of-spins 
model described above appear to be consistent with the existence of a DPT, 
although no detailed analysis has yet been made~\cite{greg3}.  At lower 
frequencies, another interesting behavior is seen in both 
the stack-of-spins and kinetic Ising models~\cite{greg3,SIDES}.  
The normalized average hysteresis-loop area,
\begin{equation}
\langle A\rangle = -\frac{1}{4M_SH_0}\oint M\left( H \right)\D H\;,
\end{equation}
is a measure of the average energy dissipation per period 
and is therefore a very important quantity.  It
is shown vs scaled frequency, $1/\varTheta$, in Fig.~\ref{areas} for the 
stack-of-spins model at $T$~$=$~$100$\,K and $T$~$=$~$20$\,K.  
At extremely low 
frequencies, the magnetization switches at very small values of 
$H$, so that $\langle A$$\rangle$~$\approx$~$0$.  At high frequencies, 
the switching rarely completes because the system is metastable 
for only a very short time interval.  Therefore, $M$ is nearly 
constant and again $\langle A\rangle$~$\approx$~$0$.  
A maximum in $\langle A\rangle$ occurs at intermediate 
frequencies $1/\varTheta$~$\approx$~$0.1$.  For studies of 
hysteresis in a kinetic Ising model which switches by a single-droplet 
mechanism, this maximum was found to correspond to stochastic 
energy resonance~\cite{SIDES}. This phenomenon has been studied further
in the kinetic Ising model~\cite{ISING4,acharyya,kim}, and also recently
investigated in models of superparamagnetic nanoparticles~\cite{raikher} and
Preisach systems~\cite{mantegna}.

\subsection{First-Order Reversal Curves}
\index{First-Order Reversal Curves}
The First-Order Reversal Curve (FORC) technique was 
developed by Pike, et al.~\cite{PIKE1} in order to 
extract more information from magnetic samples than 
is represented by, for example, the coercive field
or the remanent magnetization.  The FORC method has 
since been applied to a wide variety of systems, including several
relevant to magnetic 
nanostructures~\cite{PIKE3,carvallo,bercoff,oliva,spinu,davies,PIKE4,davies2}. 
In addition, progress has been made in understanding the role of reversible
magnetization in the FORC method \cite{PIKE5} and in improving the efficiency
of its computational use \cite{heslop}. Here, we illustrate the basic approach 
with an application to the kinetic Ising model.

The FORC technique involves decreasing the applied 
field from a positive saturating field, $H_{0}$, to a series 
of progressively more negative return fields, $H_r$, 
and recording the normalized magnetization, $m$$=$$M/M_S$, 
as the field is increased from each of these return fields 
back to the positive saturating field. This process 
results in a family of first-order reversal curves, $m(H_{r},H)$, 
where $H$ represents the applied magnetic field as 
it is increased from $H_r$ back to $H_{0}$.  Since 
the first-order reversal curves (FORCs) are determined 
by the type of reversal that has taken place before reaching $H_r$, the full 
family of FORCs should contain useful information 
about the mechanisms of reversal.

We can use the FORC method to better understand the process 
of hysteresis in the two-dimensional ferromagnetic kinetic 
Ising model on a square lattice,  choosing the Glauber 
acceptance rule to produce the dynamic of the system with 
the energy given by (10). While most FORC studies have been
done on systems with strong disorder, we focus here on the 
square-lattice Ising model without disorder.
Our simulations were performed at a temperature 
of $T=0.8~T_c$ which, given that
$k_{B}T_c \approx 2.269 J$ for the two-dimensional 
square-lattice Ising model, corresponds to 
$k_{B}T \approx 1.815 J$.  
It has been found~\cite{RIKVOLD1} that the switching of 
a fully magnetized lattice for these parameters occurs 
through single-droplet nucleation for fields up to 
$|H| \approx 0.35$, by multi-droplet nucleation 
for fields $|H| \approx$~0.35--0.9, and by strong-field 
(single-spin) reversal for fields $|H|> 0.9$.  
Since the process of switching is also influenced by 
the lattice size for finite lattices, these values serve
only as guidelines.  Here, we are mainly concerned 
with the multi-droplet regime, and so choose 
$H_{0}=0.55$.

We performed MC simulations 
to calculate the characteristic switching time $\tau$ (for 
switching from $m=1.0$ to $m<-0.8$) in a field of magnitude 
$H_0=-0.55$, finding $\tau \approx 100$ MCSS for a 
128 $\times$ 128 lattice. We therefore chose a field period of 
$P=1000$ MCSS, corresponding to a dimensionless half-period 
$\varTheta~=\frac{P/2}{ \tau} \approx 5$.  
The form of the field is taken as a sawtooth, 
piecewise linear function
\begin{equation}
{H}(t)={H}_0 \left(\frac{4|t-P/2|}{P}-1\right)\;.
\end{equation}

\begin{figure*}[t]
\vspace{0.1in}
\mbox{\subfigure[]
 {\includegraphics[width=0.49\textwidth]{figure9a.eps}}
      \quad
      \subfigure[]
 {\includegraphics[width=0.49\textwidth]{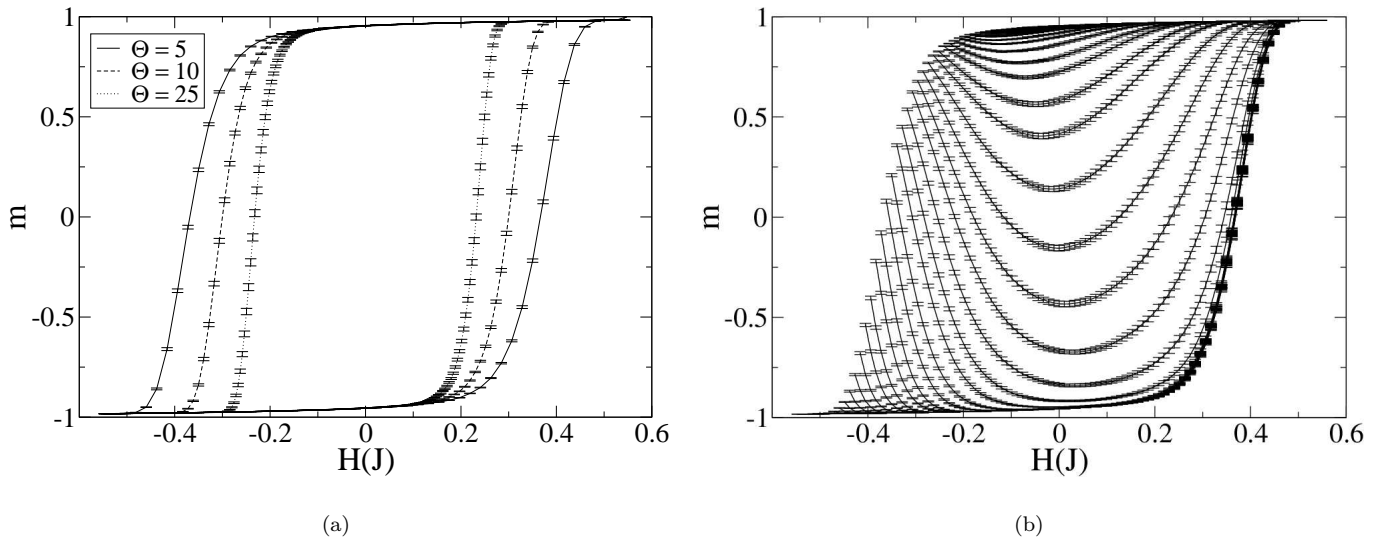}}}
\caption{({\bf a})Hysteresis loops for $\varTheta$=5 (solid line), 
         10 (dashed line), and 25 (dotted line)
         on a 128 $\times$ 128 Ising lattice. ({\bf b}) Family of FORCs 
         for the same lattice with $\varTheta$~=~5}
\label{DAN1}
\end{figure*}

Figure~\ref{DAN1}a shows the results of the 
simulation on a 128 $\times$ 128 lattice for dimensionless half-periods
of $\varTheta~=~5$, $10$, and $25$. The 
simulations were performed in parallel with 100 independent 
realizations distributed over 20 processors using the 48-bit linear 
congruential random number generator included with the 
SPRNG 2.0 package~\cite{SPRNG}.

As the lattice just completes its reversal during the full 
hysteresis loop, we expect that the family of FORCs 
will reflect much of the dynamics that are 
occurring during the reversal.  To investigate this, 
we divided the interval from $H=[-0.55,0.55]$ 
into 100 equal intervals.  We began the first FORC 
at a return field of $H_r=0.0$, and 
recorded the magnetization at $H_a$ values 
corresponding to the endpoints of the 100 intervals. 
(Thus, for the first FORC, we took 51 values of 
the magnetization.)  We then took a series of FORCs 
for $H_r$ values at the interval 
endpoints from $H=0.0$ to $H=-0.55$, 
producing a total of 51 FORCs.  For each curve, 
we averaged over 100 realizations of the MC 
simulation, a technique commonly used to find the thermally
averaged behavior of a system.
The resulting family of FORCs is shown in Fig.~\ref{DAN1}b.  
An animation of the reversal process for the FORCs shows that
the reversal does proceed by the nucleation, growth, and
shrinkage of multiple droplets (i.e., areas of reversed 
magnetization).

In a recent article \cite{robb}, we have continued this investigation
of the kinetic Ising model using the family of FORCs, as well as the FORC 
distribution, which can be derived from the FORCs as 
described in Ref.~\onlinecite{PIKE1}. The analysis yielded insights into the
limits of application of the Kolmogorov-Johnson-Mehl-Avrami (KJMA) model of 
phase transformation \cite{kjma1, kjma2, kjma3} to the
kinetic Ising system. In general, the FORC method appears to be quite
sensitive to details of the magnetization reversal process, and with some
thought can be helpful in developing insights into the construction of 
useful models.

%%%%%%%%%%%%%%%%%%%
\section{Conclusion}
%%%%%%%%%%%%%%%%%%%

Information storage devices utilizing magnetic nanostructures
have become a technologically important part of our society.
As demands for information storage increase, the size of the 
nanostructures must be decreased.  At the same time, it becomes 
important to read and write the information to these devices (i.e.
reverse the magnetization) faster.  The understanding of 
hysteresis in the magnetic nanostructures is therefore
important to the continued growth of the information-storage 
industry.  At the same time, the growth of computational
resources has provided researchers with an invaluable tool with which 
to better understand these systems.

In this overview, various common models and methods
for simulating hysteresis in magnetic nanostructures have been
presented along with results illustrating some of the properties 
of these systems.  Micromagnetic simulations are accomplished by 
integration of the Landau-Lifshitz-Gilbert (LLG) equation.  The LLG 
equation, despite being both classical and phenomenological in 
origin, nevertheless provides good insight into the magnetization 
dynamics at nanosecond time scales, provided the system is sufficiently 
finely discretized.  Our simulations on single Fe nanopillars show 
that the switching field (i.e. the field required to reduce $M_Z$ to 0)
increases continuously as the angle between the $z$-axis and 
the applied field direction is increased, consistent with experiment.
Reversal in these pillars is shown to nucleate at the endcaps 
and proceed by domain growth towards the center of the particle.  The
exception to this is the case of the applied field perpendicular to the 
long axis of the pillar, in which nucleation of reversal occurs along 
the whole length of the particle.

Unfortunately, limitations on computer resources prevent extension 
of micromagnetic simulations beyond timescales of a few tens of 
nanoseconds.  For timescales where the transition time for an individual 
spin to relax from the metastable to the stable state is much shorter than 
the time scale of interest, individual spin reversals occur with a 
probability which is related to the Boltzmann factor. The dynamics 
of the system can then be modeled using kinetic Monte Carlo 
techniques with either the Ising or Heisenberg models.  Here, we have 
shown three interesting applications of kinetic Monte Carlo simulations
of a 2-D Ising model to understanding hysteresis: dynamic phase 
transitions, stochastic resonance in the hysteresis loop area, 
and First-Order Reversal Curves (FORCs).  These illustrate only a few 
of the ways simulations of magnetic nanostructures may help give new 
insight into this important class of materials for ultra-high-density 
data storage.

\section*{Acknowledgments}
This work was supported in part by NSF grants No. DMR-0120310 and DMR-0444051,
and by the DOE Office of Science through the Computational 
Materials Science Network of BES-DMSE.

%%%%%%%%%%%%%%%%%%%%%%%% referenc.tex %%%%%%%%%%%%%%%%%%%%%%%%%%%%%%
% sample references
% "physics"
%
% Use this file as a template for your own input.
%
%%%%%%%%%%%%%%%%%%%%%%%% Springer-Verlag %%%%%%%%%%%%%%%%%%%%%%%%%%


\begin{thebibliography}{99.}

\bibitem{wikkiwikki}
http://en.wikipedia.org/wiki/Serial\_ATA

\bibitem{EWEEK}
eWeek, Jan. 17, 2006 (formerly PC Week magazine, now online only at http://www.eweek.com)

\bibitem{PLUMER} 
M.~Plumer, J.~van Ek: \emph{Micromagnetic System Modeling 
for Perpendicular Recording}, presented at the Micromagnetics 
and Magnetic Recording Workshop held at the Center for 
Materials for Information Technology, the University of Alabama 
(November 6, 2003) (unpublished)

\bibitem{GAO1}
K.Z. Gao, H.N. Bertram:
IEEE Trans. Magn. \textbf{39}, 704 (2003) 

\bibitem{GAO2}
K.Z. Gao, J. Fernandez-de-Castro, H.N. Bertram: 
IEEE Trans. Magn. \textbf{41}, 4236 (2005) 

\bibitem{DAHMEN1} 
O.~Perkovi{\' c}, K.~Dahmen, J.P.~Sethna: 
Phys. Rev. Lett. {\bf 75}, 4528 (1995)

\bibitem{DAHMEN2} 
K.~Dahmen, J.P.~Sethna: 
Phys. Rev. B {\bf 53}, 14872 (1996)

\bibitem{DAHMEN3} 
J.H.~Carpenter, K.~Dahmen, J.P.~Sethna, G.~Friedman, 
S.~Loverde, A.~Vanderveld: 
J. Appl. Phys. {\bf 89}, 6799 (2001)

\bibitem{stoner} 
E.C.~Stoner, E.P.~Wohlfarth: 
Phil. Trans. Roy. Soc. \textbf{A240}, 599 (1948)

\bibitem{garanin} 
D.A.~Garanin: 
Phys. Rev. B \textbf{55}, 3050 (1997)

\bibitem{brown} 
W.~Brown: \emph{Micromagnetics} (Wiley, 1963)

\bibitem{nowak} 
U.~Nowak: 
Thermally Activated Reversal in Magnetic Nanostructures. 
In: \emph{Annual Reviews of Computational Physics IX}, 
ed. by D.~Stauffer (World Scientific, Singapore 2001) pp. 105--152

\bibitem{greg1} 
G.~Brown, M.A.~Novotny, and P.A.~Rikvold: 
Phys. Rev. B \textbf{64}, 134432 (2001)

\bibitem{aharoni} 
A.~Aharoni:
~\emph{Introduction to the Theory of Ferromagnetism} 
(Clarendon, Oxford 1996)

\bibitem{arrot} 
A.S.~Arrot, B.~Heinrich, D.S.~Bloomberg: 
IEEE Trans. Magn. \textbf{MAG-10}, 950 (1974) 

\bibitem{CHUBYKALO} 
O.~Chubykalo, R.~Smirnov-Rueda, J.M.~Gonzalez, 
M.A.~Wongsam, R.W.~Chantrell, U.~Nowak: 
J. Magn. Magn. Mater. {\bf 266}, 28 (2003)

\bibitem{boerner2} 
E.~Boerner, H.N.~Bertram: 
IEEE Trans. Magn. \textbf{33}, 3052 (1997)

\bibitem{wirth1} 
S.~Wirth, M.~Field, D.D.~Awschalom, S.~von Moln{\'a}r: 
Phys. Rev. B \textbf{57}, R14028 (1998)

\bibitem{greengard}
L.F.~Greengard:
\emph{The Rapid Evaluation of Potential Fields in Particle Systems}
(MIT Press, Cambridge 1988)

\bibitem{greg2}
G.~Brown, T.C.~Schulthess, D.M. Apalkov, P.B. Visscher:
IEEE Trans. Magn. \textbf{40}, 2146 (2004)

\bibitem{stinnett1}
W.D.~Doyle, S.~Stinnett, C.~Dawson, L.He:
J. Magn. Soc. Jpn. \textbf{22}, 91 (1998)

\bibitem{fidler}
J.~Fidler, T.~Schrefl, W.~Scholz, D.~Suess, V.D.~Tsiantos, R.~Dittrich, M.~Kirschner:
Physica B \textbf{343}, 200 (2004)

\bibitem{hertel}
R.~Hertel, J.~Kirschner:
J. Magn. Magn. Mater. \textbf{270}, 364 (2004)

\bibitem{boerner1}
E.D.~Boerner, K.Z.~Gao, R.W.~Chantrell:
IEEE Trans. Magn. \textbf{40}, 2371 (2004)

\bibitem{He} 
L.~He, W.D.~Doyle,~H. Fujiwara: 
IEEE Trans. Magn. \textbf{30}, 4086 (1994)

\bibitem{metropolis} 
N.~Metropolis, A.W.~Rosenbluth, M.N.~Rosenbluth, 
A.H.~Teller, E.~Teller: 
J. Chem. Phys. \textbf{21}, 1087 (1953)

\bibitem{Landau}
D. Landau and K. Binder:
~\emph{A Guide to Monte Carlo Simulations in Statistical Physics} 
(Cambridge University Press, Cambridge 2000)


\bibitem{GLAUBER} 
R.J.~Glauber: J. Math. Phys. \textbf{4}, 294 (1963)

\bibitem{nowak2} 
U.~Nowak, R.W.~Chantrell, E.C.~Kennedy: 
Phys. Rev. Lett. \textbf{84}, 163 (2000)

\bibitem{chubykalo2} 
O.~Chubykalo, U.~Nowak, R.~Smirnov-Rueda, M.A.~Wongsam:
Phys. Rev. B \textbf{67}, 064422 (2003)

\bibitem{cheng1} 
X.Z.~Cheng, M.B.A.~Jalil, H.K.~Lee, Y.~Okabe:
Phys. Rev. B \textbf{72}, 094420 (2005)

\bibitem{cheng2} 
X.Z.~Cheng, M.B.A.~Jalil, H.K.~Lee, Y.~Okabe:
http://www.arxiv.org/cond-mat/0602011 (2006) (to appear in Phys. Rev. Lett.) 

\bibitem{MARTIN} 
P.~A.~Martin: 
J. Stat. Phys. \textbf{16}, 149 (1977)

\bibitem{PARK} 
K.~Park, M.A.~Novotny, P.A.~Rikvold: 
Phys. Rev. E \textbf{66}, 056101 (2002)

\bibitem{novotny} 
M.A.~Novotny: A Tutorial on Advanced Dynamic Monte Carlo Methods
for systems with Discrete State Spaces. 
In: \emph{Annual Reviews of Computational Physics IX}, 
ed. by D.~Stauffer (World Scientific, Singapore 2001) pp. 153--210
(also available at http://www.arxiv.org/cond-mat/0109182)

\bibitem{NEWPRL}
K.~Park, P.A.~Rikvold, G.M.~Buend{\'\i}a, M.A.~Novotny:
Phys. Rev. Lett., {\bf 92} 015701 (2004); 
G.M.~Buend{\'\i}a, P.A.~Rikvold, K.~Park, M.A.~Novotny:
J. Chem. Phys. \textbf{121}, 4193 (2004);
G.M.~Buend{\'\i}a, P.A.~Rikvold, M.~Kolesik:
Phys. Rev. B \textbf{73}, in press (2006)

\bibitem{bortz}
A.B.~Bortz, M.H.~Kalos, and J.L.~Lebowitz: 
J. Comput. Phys. {\bf 17}, 10 (1975)

\bibitem{MUNOZ} 
J.D.~Mu{\~n}oz, M.A.~Novotny, S.J.~Mitchell: 
Phys. Rev. E \textbf{67}, 026101 (2003)

\bibitem{watanabe} 
H.~Watanabe, S.~Yukawa, M.A.~Novotny, N.~Ito:
http://www.arxiv.org/cond-mat/0508652 (2005) (submitted to Phys. Rev. Lett.)

\bibitem{greg4} 
G.~Brown, S.M.~Stinnett, M.A.~Novotny, P.A.~Rikvold:
J. Appl. Phys. \textbf{95}, 6666 (2004)

\bibitem{wirth2}
S.~Wirth, S.~von~Moln{\'a}r:
J. Appl. Phys. \textbf{85}, 5249 (1999)

\bibitem{Li1}
Y.~Li, P.~Xiong, S.~von~Moln{\'a}r, S.~Wirth, Y.~Ohno, H.~Ohno:
Appl. Phys. Lett. \textbf{ 80}, 4644 (2002)

\bibitem{Li2}
Y.~Li, P.~Xiong, S.~von~Moln{\'a}r, Y.~Ohno, H.~Ohno:
J. Appl. Phys. \textbf{93}, 7912 (2003)

\bibitem{Li3}
Y.~Li, P.~Xiong, S.~von~Moln{\'a}r, Y.~Ohno, H.~Ohno:
Phys. Rev. B \textbf{71}, 214425 (2005)


\bibitem{TOME} 
T.~Tom\'{e}, M.J.~de~Oliveira: 
Phys. Rev. A \textbf{41}, 4251(1990)

\bibitem{MENDES} 
J.F.F.~Mendes, J.S.~Lage: 
J. Stat. Phys. \textbf{64}, 653 (1991)

\bibitem{MFA1} 
P.~Jung, G.~Gray, R.~Ray, P.~Mandel: 
Phys. Rev. Lett. \textbf{65}, 1873 (1991)

\bibitem{MFA2} 
M.F.~Zimmer: 
Phys. Rev. E \textbf{47}, 3950 (1993)

\bibitem{MFA3}
E.Z.~Meilikhov:
JETP Letters \textbf{79}, 620 (2004)

\bibitem{ISING1A} 
W.S.~Lo, R.A.~Pelcovits: 
Phys. Rev. A \textbf{42}, 7471 (1990)

\bibitem{ISING1B} 
M.~Acharyya, B.~Chakrabarti: 
Phys. Rev. B \textbf{52}, 6550 (1995)

\bibitem{ISING1C} 
M.~Acharyya: 
Phys. Rev. E \textbf{56}, 2407 (1997)

\bibitem{ISING2} 
S.W.~Sides, P.A.~Rikvold, M.A.~Novotny: 
Phys. Rev. Lett. \textbf{81}, 834 (1998);
Phys. Rev. E \textbf{59}, 2710 (1999)

\bibitem{ISING3} 
G.~Korniss, C.J.~White, P.A.~Rikvold, M.A.~Novotny: 
Phys. Rev. E \textbf{63}, 016120 (2000)

\bibitem{ISING4} 
G.~Korniss, P.A.~Rikvold, M.A.~Novotny: 
Phys. Rev. E \textbf{66}, 056127 (2002)

\bibitem{Paessens}
M. Paessens, M. Henkel:
J. Phys. A \textbf{36}, 8983 (2003)

\bibitem{TUTU} 
T.~Yasui, H.~Tutu, M.~Yamamoto, H.~Fujisaka: 
Phys. Rev. E \textbf{66}, 036123 (2002); erratum: 
\emph{ibid.} \textbf{67}, 019901(E) (2003)

\bibitem{fujiwara}
N.~Fujiwara, H.~Tutu, H.~Fujisaka:
Phys. Rev. E \textbf{70}, 066132 (2004)

\bibitem{JANG} 
H.~Jang, M.J.~Grimson, C.K.~Hall: 
Phys. Rev. B \textbf{67}, 094411 (2003)

\bibitem{huang} 
Z.G.~Huang, F.M.~Zhang, Z.G.~Chen, Y.W.~Du:
Eur. Phys. J. B \textbf{44}, 423 (2005)

\bibitem{EXP} 
Q.~Jiang, H.-N.~Yang, G.C.~Wang: 
Phys. Rev. B \textbf{52}, 14911 (1995)

\bibitem{EXP2} 
Q.~Jiang, H.-N.~Yang, G.C.~Wang: 
J. Appl. Phys. \textbf{79}, 5122 (1996)

\bibitem{FUJI} 
H.~Fujisaka,~H. Tutu,~P.A. Rikvold: 
Phys. Rev. E \textbf{63}, 016109 (2001); 
erratum: \emph{ibid.} \textbf{63}, 059903(E) (2001)

\bibitem{greg3} 
G.~Brown, M.A.~Novotny,~P.A. Rikvold: 
Physica B \textbf{306}, 117 (2001)

\bibitem{SIDES} 
S.W.~Sides, P.A.~Rikvold, M.A.~Novotny: 
Phys. Rev. E \textbf{57}, 6512 (1998)

\bibitem{acharyya}
M.~Acharyya:
Phys. Rev. E \textbf{59}, 218 (1999)

\bibitem{kim} 
B.J.~Kim, P.~Minnhagen, H.J.~Kim, M.Y.~Choi, G.S.~Jeon:
Europhys. Lett. \textbf{56}, 333 (2001)

\bibitem{raikher} 
Y.L.~Raikher, V.I.~Stepanov, R.~Perzynski:
Physica B \textbf{343}, 262 (2004)

\bibitem{mantegna} 
R.N.~Mantegna, B.~Spagnolo, L.~Testa, M.~Trapanese:
J. Appl. Phys. \textbf{97}, 10E519 (2005)

\bibitem{PIKE1} 
C.R. Pike, A. Roberts, K. Verosub: 
J. Appl. Phys. \textbf{85}, 6660 (1999)

\bibitem{PIKE2} 
C.R. Pike, A.P. Roberts, M.J. Dekkers, K. Verosub: 
Phys. Earth Planet. Inter. \textbf{126}, 11 (2001)

\bibitem{ROBERTS1} 
A. Roberts, C.R. Pike, K. Verosub: 
J. Geophys. Res. \textbf{105}, 28461 (2001)

\bibitem{PIKE3} 
C.R. Pike, A. Fernandez: 
J. Appl. Phys. \textbf{85}, 6668 (1999)

\bibitem{carvallo}
C.~Carvallo, A.R.~Muxworthy, D.J.~Dunlop, W.~Williams:
Earth Planet. Sci. Lett. \textbf{213}, 375 (2003)

\bibitem{bercoff}
P.G.~Bercoff, M.I.~Oliva, E.~Borclone, H.R.~Bertorello:
Physica B \textbf{320}, 291 (2002)

\bibitem{oliva} 
M.I.~Oliva, H.R.~Bertorello, P.G.~Bercoff:
J. Alloys Compd. \textbf{354}, 203 (2004)

\bibitem{spinu} 
L.~Spinu, A.~Stancu, C.~Radu, F.~Li, J.B.~Wiley:
IEEE Trans. Magn. \textbf{40}, 2116 (2004)

\bibitem{davies} 
J.E.~Davies, O.~Hellwig, E.E.~Fullerton, G.~Denbeaux, J.B.~Kortright, K.~Liu:
Phys. Rev. B \textbf{70}, 224434 (2004)

\bibitem{PIKE4}
C.R.~ Pike, C.A.~Ross, R.T.~Scalettar, G.T.~Zimanyi: 
Phys. Rev. B \textbf{71}, 133407 (2005)

\bibitem{davies2}
J.E.~Davies, O.~Hellwig, E.E.~Fullerton, J.S.~Jiang, S.D.~Bader, G.T.~Zimanyi, K.~Liu:
Appl. Phys. Lett. \textbf{86}, 262503 (2005)

\bibitem{PIKE5}
C.R.~ Pike: 
Phys. Rev. B \textbf{68}, 104424 (2003)

\bibitem{heslop}
D.~Heslop, A.R.~Muxworthy:
J. Magn. Magn. Mater. \textbf{288}, 155 (2005)

\bibitem{RIKVOLD1} 
P.A. Rikvold, H. Tomita, S. Miyashita, S.W. Sides: 
Phys. Rev. E \textbf{49}, 5080 (1994)

\bibitem{SPRNG} 
The SPRNG random number generator is maintained 
at Florida State University and may be downloaded 
from http://sprng.cs.fsu.edu/

\bibitem{robb} 
D.T.~Robb, M.A.~Novotny, P.A.~Rikvold:
J. Appl. Phys. \textbf{97}, 10E510 (2005)

\bibitem{kjma1} 
A.N.~Kolmogorov:
Bull. Acad. Sci. USSR, Phys. Ser. \textbf{1}, 355 (1937).

\bibitem{kjma2} 
W.A.~Johnson, R.F.~Mehl:
Trans. Am. Inst. Mining and Metallurgical Engineers \textbf{135}, 416 (1939)

\bibitem{kjma3} 
M.~Avrami:
J. Chem. Phys. \textbf{7}, 1103 (1939); \textbf{8}, 212 (1940); \textbf{9}, 177 (1941)

%
\end{thebibliography}
\end{document}